\documentclass[aps,prb,notitlepage,showpacs]{revtex4-1}

\usepackage[intlimits]{amsmath}
\usepackage{amsfonts,graphicx}

\newcommand{\bra}[1]{\left\langle #1 \right|}
\newcommand{\ket}[1]{\left| #1 \right\rangle}
\newcommand{\aver}[1]{\left\langle #1 \right\rangle}

\newcommand{\quantop}[1]{\mathcal{#1}}
\newcommand{\HH}{\quantop{H}}

\begin{document}

\title{Entanglement dynamics of second quantized quantum fields}

\author{Mikhail Erementchouk}
\author{Michael N. Leuenberger}
\affiliation{NanoScience Technology Center and Department of Physics, University of Central
Florida, Orlando, FL 32826}

\begin{abstract}
We study the entanglement dynamics in the system of coupled quantum fields. We
prove that if the coupling is linear, that is if the total Hamiltonian is a quadratic form
of field operators, entanglement can only be transferred between the fields. We show
that entanglement is produced in the model of the two-mode self-interacting boson
field with the characteristic Gaussian decay of coherence in the limit of high number
of particles. The interesting feature of this system is that the particles in different
modes become entangled even if there is no direct interaction between the modes.
We apply these results for analysis of the entanglement dynamics in the two-mode
Jaynes-Cummings model in the limit of large number of photons. While the
photon-atom interaction is assumed to conserve helicity the photons with different
polarizations still get entangled due to an effective interaction mediated by the atom
with the characteristic entanglement time linearly increasing with the number of
photons.
\end{abstract}


\maketitle

\section{Introduction}
\label{sec:intro}

Entanglement \cite{HORODECKI:2009:ID3826} is a quintessentially quantum
feature. It signifies that different parts of a compound system may form a new entity,
a complex; for example, when neither of two particles can be characterized by a
definite state so that instead of two particles one has to consider a pair and so on. If
one would attempt to access a particular part of the complex by performing a local
measurement, this would unavoidably modify the state of other parts even if the
direct interaction between the parts is absent or negligible. Such departure from the
classical properties makes entanglement the central object in various contexts, from
perspective of application in quantum informatics \cite{Bouwmeester} to
understanding physics of quantum phase transitions. \cite{WU:2004:ID4099} As a
problem of special interest, therefore, the problem of preparation a system in an
entangled state stands out. Among different appearances of this problem entangled
states of quantized electromagnetic field, as perhaps the most accessible and the
most flexible object, presents the significant importance on its own. Nowadays, the
most developed and widely used method of generating entangled photons is the
parametric down conversion,\cite{Shih,Ou} which is based on the two-photon
radiative decay of material excited states. This method, however, suffers from well
recognized intrinsic limitations --- very low yield and rescaling the wavelength of the
emitted photons.\cite{Bouwmeester,Mandel,Tanzilli} Therefore, there is the constant
search of alternative sources of entangled
light,\cite{Kwiat,EDAMATSU:2004:ID3421,Altepeter,STEVENSON:2006:ID3501,Young,Akopian,Shields,Hayat2007,Oohata,Hayat2008,EREMENTCHOUK:2010:ID4065}
which motivates a thorough consideration of entanglement of quantum many-body
states.

The characteristic feature of the process of entangling photons in the course of
interaction with matter is nonconserving number of photons, or involved particles in
general. In the few particle limit or in the case when the typical time scales are well
separated, these processes can be reduced to the more or less standard quantum
mechanical situation. This makes the process of entangling fitted into the well
developed description of entanglement. Indeed, in this case one can specify time
periods when the system is either in the state of excited matter and no (relevant)
photons, or in the state when matter is in the ground state and there are emitted
photons. The general problem of solid based sources of entangled light, however,
compels addressing more general situation, when the photon states coexist with
material excitations and the processes of reabsorption and re-emission may play an
important role. In this case one has to incorporate nonconserving number of particles
fully into consideration. The quantum field description (more precisely, the formalism
of second quantization) provides the most natural framework for dealing with this
kind of situation. In this approach particles appear not as predefined entities as, say,
qubits within the standard quantum mechanical treatment, but rather as excitations
of the respective quantized fields.

As we will demonstrate below, entanglement in the context of
quantized fields has its subtleties. The complete description
of entanglement in this context is yet to be developed. For our
purposes, however, a basic approach is sufficient. First, we
will restrict our attention only to the case when the whole
system can be characterized by a pure state. Second,
entanglement will be considered from the perspective of the
problem of sources of entangled light, which advances
consideration of particle properties, as will be elaborated
below. From this point of view the existence of entanglement
implies that a set of particles is in a nonseparable state and,
as a result, one particle cannot be described by a state
vector, the particle with necessity is in a mixed state. Pure
and mixed states, in turn, can be distinguished using the fact
that extremal values of observables corresponding to operators
with non-degenerate spectrum are reached at pure states. Let
$\widehat{O}$ be an operator acting on a finite dimensional
Hilbert space and let its spectral decomposition be
$\widehat{O} = \sum_\kappa O_\kappa \widehat{\Pi}_\kappa$,
where $O_\kappa$ are the eigenvalues, enumerated in the
ascending order, $O_1 < O_2 \leq O_3 \ldots \leq O_N$, and
$\widehat{\Pi}_\kappa$ are the projectors on the respective
eigen-spaces. Furthermore, let the smallest eigenvalue be
non-degenerate, i.e. $\mathrm{rank}(\widehat{\Pi}_1) = 1$, then
the minimal value of $\aver{\widehat{O}} \equiv
\mathrm{Tr}\left[\widehat{\rho}\widehat{O}\right]$ (understood
as a function of state described by the density matrix
$\widehat{\rho}$) is equal to $O_1$ and is reached at
$\widehat{\rho}$ corresponding to the pure state,
$\widehat{\rho} = \widehat{\Pi}_1$. Respectively, the value of
the observable $\aver{\widehat{O}}$ on mixed states will always
satisfy inequality $\aver{\widehat{O}}_{mix} > O_1$. Indeed, in
the basis, where $\widehat{O}$ is diagonal, one has
$\mathrm{Tr}\left[\widehat{\rho}\widehat{O}\right] =
\sum_\kappa O_\kappa \rho_{\kappa, \kappa} \geq O_1 \sum_\kappa
\rho_{\kappa, \kappa} = O_1$ with the equality reached only
when all $\rho_{\kappa, \kappa}$ but $\rho_{1,1}$ are zero.
Thus, in a sense any operator with nondegenerate extremal
eigenvalues allows distinguishing between pure and mixed
states. Using this observation one can implement the ideology
of witnessing entanglement, which originally was developed in
the context of the full compound system
\cite{HORODECKI:2009:ID3826,VEDRAL:2008:ID4032} and later was
extended to one particle description. \cite{TERRA:2007:ID4022}

This approach can be directly applied to the quantum field description. In this case
the role of witnesses is played by the respective one-particle operators and following
the standard line of arguments (see e.g. Ref.~\onlinecite{ZAGOSKIN:1998:ID3943})
one can characterize entanglement of one particle with the rest of the system using
one particle correlation matrix (OPCM) $G_{\kappa, \lambda} =
\aver{a^\dagger_\kappa a_\lambda}$, where $\kappa$ and $\lambda$ enumerate
one-particle states and $a^\dagger_\kappa$ and $a_\lambda$ create and destroy a
particle in the respective states. Throughout the paper we incorporate the time
dependence into the Heisenber representation of the operators $a_\kappa(t) = \exp(i
\HH t) a_\kappa \exp(-i \HH t)$, where $\HH$ is the Hamiltonian describing the
whole system. Thus, the time dependence of the OPCM is given by
\begin{equation}\label{eq:SDM_def}
  G_{\kappa, \lambda}(t) = \aver{a^\dagger_\kappa (t) a_\lambda(t)},
\end{equation}
where the average is taken with respect to the initial state $\aver{\ldots} =
\bra{\psi(t=0)}\ldots \ket{\psi(t=0)}$.

In order to illustrate the difficulty of producing entangled states let us consider a
simple example of a boson field driven by an external source
\begin{equation}\label{eq:boson_classic}
  \quantop{H} = \sum_{\kappa} \left(\epsilon_\kappa a^\dagger_\kappa a_\kappa +
  e_\kappa(t)a^\dagger_\kappa + e_\kappa^*(t) a_\kappa\right),
\end{equation}
where $e_\kappa(t)$ are $c$-numbers determined by the projections of, generally
speaking time dependent, external classical field onto respective one-particle states.

Before applying rigorous methods let us note that this system may look confusing if
approached with the help of often employed arguments based on interference of
different paths connecting the initial and final states.  Indeed, considering that there
are different ways to fill some particular state, say, with only two particles, one might
expect that these two particles will become entangled, which, of course, is an
incorrect conclusion.

In virtue of the discussion above, the absence of entanglement would be manifested
by rank one of the OPCM,\cite{PASKAUSKAS:2001:ID3811,WANG:2005:ID3868}
while the rank of the OPCM can be easily investigated. The solutions of the operator
equations of motion have the form
\begin{equation}\label{eq:classic_boson_sol}
  a_\kappa(t) = a_\kappa(0)\exp\left(-i \epsilon_\kappa t\right) + E_\kappa(t),
\end{equation}
where $E_{\kappa}(t) = \int_0^t dt' \exp[-i\epsilon_\kappa(t-t')]e_\kappa^*(t')$
are $c$-number functions. Substituting this representation into
Eq.~\eqref{eq:SDM_def} we find
\begin{equation}\label{eq:classic_SDM_time}
 G_{\kappa, \lambda}(t) = e^{it (\epsilon_\kappa - \epsilon_\lambda)}G_{\kappa, \lambda}(0) +
 E_\kappa^*(t) \aver{a_\lambda} e^{-i \epsilon_\lambda t}
 + \aver{a_\kappa^\dagger} E_\lambda(t)  e^{i \epsilon_\kappa t} +
 E_{\kappa, \lambda}(t),
\end{equation}
where $E_{\kappa, \lambda}(t) = E^*_\kappa(t)E_\lambda(t)$. Generally the
structure of the OPCM $\widehat{G}(t)$ (here and in the following hats denote
matrices in the space of one-particle states) driven by the external source depends on
time nontrivially. The last three terms in Eq.~\eqref{eq:classic_SDM_time} may lead
to variation of entanglement depending on the structure of the initial state. If,
however, the system is initially in the vacuum state, i.e. $\widehat{G}(0) = 0$, then
$\widehat{G}(t) = \widehat{E}(t)$, where $\widehat{E}(t)$ is the matrix with the
elements $E_{kq}(t)$. In turn $\widehat{E}(t)$ is at most of rank one implying the
absence of entanglement.


This example shows that the problem of the dynamics of entanglement should be
treated with certain care. First, the naive arguments based on the picture of
interference of different paths may be misleading. Second, no matter how complex is
the internal dynamics of the system described by the spectrum, $\epsilon_\kappa$,
and independently on particular time dependence of the external excitation
$e_\kappa(t)$, the states, reached out of vacuum under the action of this excitation,
are disentangled. In particular, the initial state, which may lead to a nontrivial time
dependence of entanglement, as has been mentioned above, must be created by
other means than the external classical excitation.

The rest of the paper is organized as follows. In Section~\ref{sec:pictures} we
provide more detailed description of entanglement in the context of quantized fields.
In Section~\ref{sec:coupled} we consider the entanglement dynamics for the system
of linearly coupled boson fields. In Section~\ref{sec:bosons} we study entanglement
for the case of two-mode self-interacting boson field. Finally, in
Section~\ref{sec:jc-model} we apply the obtained results to the analysis of the
entanglement dynamics for the two-mode Jaynes-Cummings model.

\section{Entanglement within field and particle pictures}
\label{sec:pictures}

Some results presented in Sections~\ref{sec:coupled} and \ref{sec:bosons} may
seem to contradict results readily available in the literature. This reflects a certain
ambiguity of the notion of entanglement in the context of quantized fields. Therefore,
in order to avoid possible misunderstanding, it is useful to analyze the problem of
entanglement in details.

The ambiguity stems from the fact that entanglement is understood as a relation
between a part of the compound system and the whole system, while there are two
distinctive notions of the part when quantized fields are considered, fields and
particles. Which of these two different entities, fields and particles, appear more
naturally is dictated by the physical content of the specific problem. For example, if a
system of harmonic or unharmonic oscillators is considered
\cite{CHUNG:2009:ID4045}, individual oscillators (i.e. fields) stand out as the part of
the big system, while, say, dynamics of excitons in semiconductors
\cite{EDAMATSU:2004:ID3421,STEVENSON:2006:ID3501,Akopian,Hayat2007,Oohata,EREMENTCHOUK:2010:ID4065}
promotes consideration of particles.

An arbitrary (pure) state of a system of $M$ (boson) fields can be specified in terms
of degrees of excitation (population numbers) of each field
\begin{equation}\label{eq:state_field_p}
  | \psi \rangle = \sum_{n_1, \ldots, n_M}  \Psi_{n_1, \ldots, n_N} |n_1, \ldots, n_M \rangle,
\end{equation}
where $n_\kappa$ denotes the population of the $\kappa$-the field, and $\Psi_{n_1,
\ldots, n_M}$ are the respective amplitudes. Equivalently, this state can be presented
in terms of the particle creation operators
\begin{equation}\label{eq:state_particle_p}
  | \psi \rangle = \sum_{N=0}^\infty \sum_{\kappa_1, \ldots, \kappa_N}
  \Phi_{\kappa_1, \ldots, \kappa_N} a_{\kappa_1}^\dagger \ldots a_{\kappa_N}^\dagger |0 \rangle,
\end{equation}
where $|0\rangle$ is vacuum, $N$ is the total number of particles and
$a_\kappa^\dagger$ creates a particle in the $\kappa$-the one-particle state.
Representations \eqref{eq:state_field_p} and \eqref{eq:state_particle_p} are
equivalent if the amplitudes $\Phi_{\kappa_1, \ldots, \kappa_N}$ satisfy conditions
that follow from the commutation of the creation operators $\left[a_\kappa^\dagger,
a_\lambda^\dagger\right]=0$ and the relation between the individual fields in
Eq.~\eqref{eq:state_field_p} and one-particle states in
Eq.~\eqref{eq:state_particle_p}, $|n \rangle = \left(a^\dagger\right)^n|0 \rangle
/\sqrt{n!}$.

While representations \eqref{eq:state_field_p} and \eqref{eq:state_particle_p} are
equivalent they imply different notions of ``locality''. This follows simply from the
following observation. An operator acting on a particular \textit{field} in
Eq.~\eqref{eq:state_field_p} does not preserve the total number of particles unless
it's proportional to identity operator. At the same time an operator, which changes the
state of one \textit{particle} in Eq.~\eqref{eq:state_particle_p}, obviously affects
the population numbers of more than one field. In turn, entanglement strongly relies
on the notion of locality and, therefore, is sensitive to the choice of eligible local
transformations. This compels to draw a distinction between representations
\eqref{eq:state_field_p} and \eqref{eq:state_particle_p}. For this reason we will call
representation~\eqref{eq:state_field_p} \textit{the field picture} and
Eq.~\eqref{eq:state_particle_p} will be referred to as \textit{the particle picture}.
Most of the papers dealing with entanglement in the field context are effectively
restricted to one of the pictures implied by the physical situation. This, however, leads
to possible ambiguities because results are often formulated in some general terms
(entanglement, part, compound system and so on), which are identical for both
pictures. Entanglement, however, does depend on the picture, as is suggested by the
nonequivalence of the notions of locality and will be demonstrated below.

Within the field picture, if the general form of operators acting on a particular field is
allowed, entanglement is naturally related to separability of amplitudes $\Psi_{n_1,
\ldots, n_M}$. In systems with superselection rules\cite{BARLETT:2003:ID3996} the
space of allowed operators is ``smaller" and the operators, whose mean value is
extremal at the particular state, may not be accessible. From the perspective of
witness ideology this means that such state should be considered as entangled. For
such systems, therefore, one needs criteria of entanglement different from mere
separability of the respective amplitudes. This problem, however, goes beyond the
scope of the present paper.

In order to quantify entanglement of the $\kappa$-th field with the rest of the
system, i.e. with the other fields,  it is convenient to introduce the reduced density
matrix
\begin{equation}\label{eq:reduced_DM_field}
  \rho_{n,n'}^{(\kappa)} = \sum_{n_1,\ldots, n_N} \langle n_1, \dots, n_{\kappa-1}, n, \dots, n_M | \psi \rangle
  \langle \psi | n_1, \dots, n_{\kappa-1}, n', \dots, n_M \rangle,
\end{equation}
where the summation excludes $n_\kappa$. If the rank of the reduced density matrix
$\rho_{n,n'}^{(\kappa)}$ is higher than one, we have entanglement, which can be
quantified, for example, by the von Neumann entropy. Thus, entanglement in the
field picture fits the canonical quantum mechanical description based on separability
of the amplitudes and agrees with the general witness ideology.

Within the particle picture, however, one has to rely upon the witness approach
heavier due to indistinguishability of individual particles, which makes the separability
property of amplitudes $\Phi_{\kappa_1, \ldots, \kappa_N}$ an inadequate
criterium. The quantity of main interest becomes the one-particle correlation matrix
\begin{equation}\label{eq:opcm_def}
  G_{\kappa,\lambda} = \aver{a^\dagger_\kappa a_\lambda},
\end{equation}
which is related to the amplitudes $\Phi_{\kappa_1, \ldots, \kappa_N}$ in rather
complex way. The one-particle density matrix (OPDM) is defined as properly
normalized correlation matrix $\widehat{\rho} = \widehat{G}/ \mathrm{Tr}
[\widehat{G}]$. After the normalization one can quantify entanglement in the
particle picture using von Neumann entropy of OPDM
\begin{equation}\label{eq:ent_opdm}
  E_N[\widehat{\rho}] = - \mathrm{Tr}\left[\widehat{\rho} \log(\widehat{\rho})\right].
\end{equation}
One can easily check that when the system immediately admits
the standard description (e.g. when all particles are
distinguishable, i.e. all $\kappa_i$ in
Eq.~\eqref{eq:state_particle_p} are different and the field and
particle pictures are identical) the approach based on
Eqs.~\eqref{eq:opcm_def} and \eqref{eq:ent_opdm} yields results
consistent with this description. Of course, OPDM may only
answer questions regarding entanglement of a single particle
with the rest of the system. If one is interested in more
subtle details, such as, for instance, entanglement in pairs,
one has to look at the density matrices of higher order, say,
two-particle density matrices for pairs. We, however, limit
ourselves to studying the basic properties of entanglement and
for this purpose it suffices to consider OPDM, which can be
shown to yield the upper bound for entanglement in the system.

The necessity to distinguish entanglement within different pictures is illustrated by
the states with amplitudes (in the field picture) being separable
\begin{equation}\label{eq:sep_amp_tricky}
  \Psi_{n_1,\ldots, n_N} = \prod_{i=\kappa}^M \psi^{(\kappa)}_{n_\kappa}
\end{equation}
and for each field only amplitudes of the same parity are not zero, for example,
$\psi^{(\kappa)}_{2n} = 0$ for all $\kappa$ and $n \geq 0$. These states are
disentangled within the field picture; they, however, may be completely entangled
within the particle picture. Indeed, the off-diagonal elements of the OPDM vanish for
such states and if, additionally, the amplitudes $\psi^{(\kappa)}_{2n+1}$ are
chosen in such manner that the average number of particles in each state (the degree
of excitation of each field) is the same, $\aver{a_\kappa^\dagger a_\kappa} =
\mathrm{const}$, then the von Neumann entropy takes the maximal value implying
maximal entanglement within the particle picture.

The example of opposite situations is presented by single-particle entanglement,
\cite{TERRA:2007:ID4022} when, say, $\Psi_{1,0}\Psi_{0,1} \ne 0$ while all other
amplitudes in Eq.~\eqref{eq:state_field_p} are zero. These states are obviously
entangled within the field picture while disentangled in the framework of the particle
picture.

For more detailed comparison of different pictures we show that the states
disentangled within both, field and particle, pictures constitute the special class and
this class is not particularly rich. Besides the states with only excitations of one field,
when trivially there is nothing to entangle (we would like to remind that we consider
the case without superselection rules), these are canonical coherent
states.\cite{PERELOMOV:1986:ID4096,GAZEAU:2009:ID4097,ZHANG:1990:ID4095}

First of all we describe all disentangled states within the particle picture. The main
result here is almost obvious: \textit{particles are disentangled within the particle
picture if and only if they all are in the same state}. In order to prove this statement
(nontrivial in the part that there are no other disentangled states) it is convenient to
introduce special notations. In the context of the particle picture operators
$a_\kappa^\dagger$ and $a_\kappa$ create and destroy particles in the respective
one-particle states enumerated by the index $\kappa$, which constitute the
one-particle Hilbert state $H$. For simplicity we assume that the dimension of $H$ is
finite, $M$. This assumption is not crucial for the following consideration while allows
us to avoid some formal complications.

We introduce a vector-operator $\mathbf{a}$ with components
$(\mathbf{a})_\kappa = a_\kappa$. The commutation relation in terms of the
vector-operators can be formally written as $\left[\mathbf{a}^\dagger,
\mathbf{a}\right] = - \widehat{1}$. These are vectors in the following sense. The
choice of a different set of basis one-particle states in space $H$ corresponds to a
linear transformation of $H$, which translates into choosing a different set of
operators $b_\lambda$ linearly related to the old set
\begin{equation}\label{eq:transformation}
b_\lambda = \sum_\kappa a_\kappa U_{\kappa, \lambda}
\end{equation}
or $\mathbf{b} = \mathbf{a} \widehat{U}$. The commutation relation for new
operators can be shown to have the form $\left[\mathbf{b}^\dagger,
\mathbf{b}\right] = - \widehat{U}^\dagger \widehat{U}$. Thus, in order to satisfy
the boson commutation relation the vector-operators $\mathbf{b}$ and
$\mathbf{a}$ must be related through a unitary transformation. These
transformations, in particular, preserve the operator of the total number of particles.
This can be illustrated presenting the operator as $\mathcal{N} = \sum_\kappa
a^\dagger_\kappa a_\kappa \equiv \mathrm{Tr}\left(\mathbf{a}^\dagger \otimes
\mathbf{a}\right)$. Here and below in this section $\otimes$ denotes the product
$\left(\mathbf{a}^\dagger \otimes \mathbf{a}\right)_{\kappa, \lambda} =
a^\dagger_\kappa a_\lambda$, which transforms as a tensor.

The reason why we have introduced vector-operators is that OPCM also transforms as
a tensor
\begin{equation}\label{eq:OPCM_transformation}
  \aver{\mathbf{b}^\dagger \otimes \mathbf{b}} = \widehat{U}^\dagger
  \aver{\mathbf{a}^\dagger \otimes \mathbf{a}} \widehat{U},
\end{equation}
i.e. $\widehat{G}$ is mapped into $\widehat{U}^\dagger
\widehat{G}\widehat{U}$. Thus, unitary transformations do not change the
spectrum of OPCM and, due to invariance of $\mathcal{N}$, they leave entanglement
intact.

The important consequence of this geometrical picture is that it immediately provides
the description of all disentangled states.  OPCM is a Hermitian matrix and, therefore,
can be diagonalized by a unitary transformation implying the existence of the
preferred set of operators. In turn, OPCM of \textit{any} disentangled state in the
diagonal form has only single non-zero element. Thus, choosing the appropriate
transformation we can have only $\aver{b^\dagger_1 b_1} \ne 0$ while all other
elements of OPCM are zero. The only states yielding such OPCM are of the form
\begin{equation}\label{eq:disent_b}
  \ket{\psi} = \sum_n \frac{\phi_n}{\sqrt{n!}} \left(b_1^\dagger\right)^n \ket{0},
\end{equation}
where $\sum_n |\phi_n|^2 = 1$. Using relation~\eqref{eq:transformation} we can
expand $ b_1^\dagger = \sum_\kappa U^*_{\kappa,1} a_\kappa^\dagger$. As
follows from unitarity the single column of a unitary matrix is a unit vector. Denoting
this vector by $\mathbf{S}$ and defining the ``scalar" product $\mathbf{S} \cdot
\mathbf{a}^\dagger \equiv \sum_\kappa S_\kappa a_\kappa^\dagger$ we can
parametrize all disentangled states
\begin{equation}\label{eq:disent_general}
 \ket{\psi_\mathbf{S}(\phi_1, \ldots)} = \sum_n \frac{\phi_n}{\sqrt{n!}} \left(\mathbf{S} \cdot \mathbf{a}^\dagger\right)^n \ket{0}
\end{equation}
by a set of amplitudes $\phi_n$ and a vector on the unit sphere in $\mathbb{C}^M$
(or in the one-particle Hilbert state).

Once we've established the general form of disentangled states in the particle picture
we may proceed and find which of those are disentangled in the field picture. In order
to find amplitudes $\Phi_{n_1,\ldots , n_M} = \aver{n_1, \ldots, n_M |
\psi_\mathbf{S}}$ we first notice that only the term with the same total number of
particles $n = \sum_i n_i$ in Eq.~\eqref{eq:disent_general} into $\Phi_{n_1,\ldots ,
n_M}$ and from the polynomial expansion of this term we need only one term with
matching population numbers for each field. Thus we find
\begin{equation}\label{eq:amplitudes_disent}
 \Phi_{n_1,\ldots , n_2} = \phi_n \sqrt{n!} \prod_\kappa S_\kappa^{n_\kappa} \sqrt{n_\kappa!}
\end{equation}
with $n=\sum_\kappa n_\kappa$ and the important convention that $S_\kappa^0 =
1$ even if $S_\kappa = 0$. It follows from Eq.~\eqref{eq:amplitudes_disent} that
there are only two possibilities to have separability of $\Phi_{n_1,\ldots , n_M}$. The
first one is when all but one $S_\kappa$ are zero. This corresponds to the trivial case
when only the single type of fields is excited. The second possibility is when $\phi_n
= c \alpha^n/\sqrt{n!}$ with some complex numbers   $c$ and $\alpha$.
Substituting these amplitudes into Eq.~\eqref{eq:disent_general} and enforcing the
normalization condition we find that all states disentangled in both, field and particle,
pictures can be presented as
\begin{equation}\label{eq:coherent_disentangled}
 \ket{\psi_\mathbf{S}(\alpha)}  = \exp\left(\alpha \, \mathbf{S}\cdot \mathbf{a}^\dagger - \alpha^* \mathbf{S}^*\cdot \mathbf{a}\right)
 \ket{0}.
\end{equation}
These are canonical coherent states: they satisfy the equation $\mathbf{S}^*\cdot
\mathbf{a}\ket{\psi_\mathbf{S}(\alpha)} = \alpha
\ket{\psi_\mathbf{S}(\alpha)}$. It is interesting to emphasize a relation with the
example considered in the Introduction. Hamiltonian~\eqref{eq:boson_classic} is
diagonalized introducing operators $b_\kappa = a_\kappa +
e_\kappa/\epsilon_\kappa$, which are obtained by employing Glauber's shift operator
$\exp(\alpha \mathbf{S}\cdot \mathbf{a}^\dagger - \alpha^* \mathbf{S}^*\cdot
\mathbf{a})$ with $\alpha S_\kappa = e_\kappa/\epsilon_\kappa$. This, in
particular, proves that states reached from vacuum under classical excitation are
disentangled within both pictures.

The general ``inverse" problem of relation between entanglement and the structure
of respective states does not succumb to such simple analysis and requires
consideration, which goes far beyond the objectives of the present publication.
Therefore we limit ourselves to explicit description of a few sets of completely
entangled states.

In Section~\ref{sec:bosons} states of NOON type appear. While these states were
introduced for qubits \cite{BOTO:2000:ID4100} they can be defined in a more
general setup as follows. Let vectors $\mathbf{S}(\kappa)$, $\kappa = 1, \ldots, M$
form a basis in $\mathbb{C}^M$, then NOON states of $N$ particles with $M$
dimensional one-particle Hilbert space is
\begin{equation}\label{eq:NOON_def}
  \ket{NOON} = \frac{1}{\sqrt{MN!}} \sum_\kappa e^{i \chi_\kappa}\left[\mathbf{S}(\kappa) \cdot \mathbf{a}^\dagger \right]^N \ket{0},
\end{equation}
where we have added phase factors $e^{i \chi_\kappa}$ in order to allow for sign
variation between different terms in the sum. These factors, however, can be
incorporated into the basis vectors $\mathbf{S}(\kappa) \to \mathbf{S}(\kappa)
e^{i \chi_\kappa/N}$. Thus, taking account that different rearrangements of
$\mathbf{S}(\kappa)$ produce the same state,  the manifold of NOON states is
isomorphic to $SU(M)/S_M$, where $S_M$ is the symmetric group (group of all
permutations of $M$ elements).

It should be noted that NOON states do not exhaust all completely entangled states.
In order to see this it is useful to present NOON states as they appear in a more
general context. The natural representation of $S_M$ on the basis
$\mathbf{S}(\kappa)$ has the form $T(g) \mathbf{S}(\kappa) =
\mathbf{S}(g^{-1}\kappa)$ for any $g\in S_M$. Taking any $1 \leq \kappa \leq M$
NOON states can be defined in terms of the orbits
\begin{equation}\label{eq:NOON_def_orbit}
  \ket{NOON} = \frac{\sqrt{M}}{M! \sqrt{N!}} \sum_{g\in S_M} \left[\mathbf{S}(g^{-1}\kappa)\cdot \mathbf{a}^\dagger\right]^N \ket{0},
\end{equation}
where we have taken into account that the orbit $g^{-1}\kappa$ on
$\{1,\ldots,M\}$ visits each element $(M-1)!$ times as $g$ runs over $S_M$. Using
the same approach another set of completely entangled states can be constructed for
$N>3$. For any pair $1\leq \kappa, \lambda \leq M$ such that $\lambda \ne \kappa$
states
\begin{equation}\label{eq:n11n_def}
  \ket{N-1,1} = \frac{\sqrt{M}}{M! \sqrt{(N-1)!}} \sum_{g\in S_M} \left[\mathbf{S}(g^{-1}\kappa)\cdot \mathbf{a}^\dagger\right]^{N-1}
  \mathbf{S}(g^{-1}\lambda)\cdot \mathbf{a}^\dagger \ket{0}
\end{equation}
are completely entangled. This makes an interesting connection between completely
entangled states (constituting a straightforward generalization of Dicke states
\cite{MAKELA:2007:ID4094,MARKHAM:2011:ID4093}) and irreducible
representations of symmetric group $S_N$.

These observations suggest that the natural framework for dealing with entanglement
in particle picture is provided by the language of irreducible representations of
$SU(M)$ Lie groups. For example, in the case of two-dimensional one-particle Hilbert
space (e.g. photons characterized by two ``$+$" and ``$-$" polarizations) it is
convenient to employ Schwinger's model of angular
momentum\cite{SAKURAI:1994:ID3942} and to introduce
\begin{equation}\label{eq:ang_momentum_ops}
\begin{split}
  \quantop{J}_x = \frac 1 2 (a_+^\dagger a_- + a_-^\dagger a_+), \\
  \quantop{J}_y = \frac 1 {2i} (a_+^\dagger a_- - a_-^\dagger a_+), \\
  \quantop{J}_z = \frac 1 2 (a_+^\dagger a_+ - a_-^\dagger a_-),
\end{split}
\end{equation}
which satisfy the commutation relation of $\mathfrak{su}(2)$ algebra
$\left[\quantop{J}_\kappa, \quantop{J}_\lambda\right] = i \epsilon_{\kappa,
\lambda, \mu} \quantop{J}_\mu$ with $\epsilon_{\kappa, \lambda, \mu}$ being
completely antisymmetric tensor. Together with operator $\quantop{J}_0 =
(a_+^\dagger a_+ + a_-^\dagger a_-)/2$ these operators provide a representation
of OPDM
\begin{equation}\label{eq:DM_algebra}
  \widehat{\rho} = \sum_{i=0}^3 \sigma_i \aver{\quantop{J}_i},
\end{equation}
where $\sigma_i$ with $i=1,2,3$ are Pauli matrices and $\sigma_0$ is the identity
matrix. As we will show below, in this case entanglement is unambiguously expressed
in terms of $\sum_{i=1}^3 \aver{\quantop{J}_i}^2 = J^2$, that is a quantity
determined completely by the generators of the Lie algebra.

\section{Entanglement transfer between coupled fields}
\label{sec:coupled}

We begin our analysis from considering the dynamics of entanglement for a simple
but important case of linearly coupled fields described by the
Hamiltonian
\begin{equation}\label{eq:coupled_H}
  \quantop{H} = \sum_k \epsilon^{(a)}_k a_k^\dagger a_k +
  \sum_{\kappa} \epsilon^{(b)}_{\kappa} b_{\kappa}^\dagger b_{\kappa}
  + \sum_{k,\kappa} \left( d_{k,\kappa} a_k^\dagger b_\kappa +
  d_{k,\kappa}^* b_\kappa^\dagger a_k \right),
\end{equation}
where $k$ and $\kappa$ enumerate the modes of the fields,
$\epsilon^{(a)}_k$, $\epsilon^{(b)}_\kappa$ and $d_{k,\kappa}$ are the
spectra of the fields and the coupling constants between them,
respectively, and the operators $a_k$ and $b_\kappa$ are assumed to obey
the boson commutation relations.

The important feature of the time evolution of entanglement in this system is that the
total entanglement remains constant and is solely determined by the initial state. For
more precise formulation, instead of the operators $a_k$ and $b_\kappa$ let us
introduce the combined operators $u_i$. That is instead of two fields with two sets of
modes $M_a$ and $M_b$ we consider the single field, whose modes are the direct
sum $M_a \oplus M_b$. Thus $u_i \equiv a_{k_i}$ if $i \in M_a$ and $u_i \equiv
b_{\kappa_i}$ if $i \in M_b$. For field $u$ we define the OPCM
\begin{equation}\label{eq:b_spdm}
  G_{ij}(t) = \aver{u_i^\dagger(t) u_j(t)}.
\end{equation}
When both $i$ and $j$ belong to, say, $M_a$ the respective matrix elements
$K_{ij}(t)$ give the OPCM for field $a$ and so on. Furthermore, let the matrix
$\widehat{K}$ have the spectral representation
\begin{equation}\label{eq:u_spdm_spectral}
\widehat{G}(t) = \sum_{\mathfrak{l}} \lambda_{\mathfrak{l}}(t)
 \mathbf{v}_{\mathfrak{l}}(t) \otimes \mathbf{v}_{\mathfrak{l}}(t),
\end{equation}
where $\lambda_{\mathfrak{l}}$ and $\mathbf{v}_{\mathfrak{l}}$ are the
eigenvalues and the unit eigenvectors of $\widehat{K}$, respectively, and $\otimes$
denotes the tensor product, which is defined as $\left(\mathbf{v}\otimes
\mathbf{v}'\right)_{ij} = {v}^*_i v'_j$. Then the total entanglement is defined as
the von Neumann entropy
\begin{equation}\label{eq:u_ent}
  E_N(t) = - \sum_{\mathfrak{l}} \widetilde{\lambda}_{\mathfrak{l}}(t)
  \log\left[\widetilde{\lambda}_{\mathfrak{l}}(t)\right],
\end{equation}
where $\widetilde{\lambda}_{\mathfrak{l}}(t) =
\lambda_{\mathfrak{l}}(t)/\sum_{\mathfrak{m}} \lambda_{\mathfrak{m}}(t)$.

The important feature of systems with linear coupling between fields is that the total
entanglement remains constant and is determined by the initial state $E_N(t) =
E_N(0)$. Therefore, the entanglement evolution restricts purely to its redistribution
between the fields. This result while specific for the particle picture (thus contrasting
the results obtained, e.g. in
Refs.~\onlinecite{VEDRAL:2003:ID3974,CHUNG:2009:ID4045}) holds in a more
general context agreeable with the description of entanglement in terms of irreducible
representations of $SU(M)$ Lie groups. Therefore, we prove it for a system described
by the Hamiltonian
\begin{equation}\label{eq:H_general}
  \HH = \sum_{ij} h_{ij} u^\dagger_i u_j,
\end{equation}
with hermitian $\widehat{h}$ and \textit{pairs} of field
operators obeying the commutation relation
\begin{equation}\label{eq:u_cr}
  [u^\dagger_i u_j, u^\dagger_k u_l] = u^\dagger_i u_l \delta_{kj} - u^\dagger_k u_j \delta_{il}.
\end{equation}
The total entanglement is again the von Neumann entropy of the OPCM, $G_{ij}(t)$,
which satisfies
\begin{equation}\label{eq:ug_eq_motion}
  \frac{\partial }{\partial t} \widehat{G}(t) = i\left[\widehat{h}, \widehat{G}\right].
\end{equation}
The spectrum of matrices, whose time dependence is governed by
such equations with the commutator in the r.h.s., does not
change with time. Indeed, the solution of
Eq.~\eqref{eq:ug_eq_motion} has the form $\widehat{G}(t) =
\exp(i \widehat{h}t)\widehat{G}(0) \exp(-i \widehat{h}t)$. Thus
due to unitarity of $\exp(i \widehat{h}t)$ the spectral
representation of $\widehat{G}(t)$ is given by
Eq.~\eqref{eq:u_spdm_spectral} with constant
$\lambda_\mathfrak{l}$ and only the eigenvectors are functions
of time. This implies that entanglement is an integral of
motion.

For example, if initially there was only single non-zero
eigenvalue (i.e. entanglement was zero) it remains the only one
later on implying no production of entanglement. The same
result holds for the OPCM corresponding to either fields $a$ or
$b$. Indeed, the OPCM of the field $a$ is obtained from the
OPCM of the combined field $u$ applying the respective
projection operators, $\widehat{\Pi}_a$, so that
\begin{equation}\label{eq:Ka_from_Ku}
  \widehat{G}^{(a)} = \widehat{\Pi}_a \widehat{G} \widehat{\Pi}_a.
\end{equation}
Using for $\widehat{G}$ its spectral representation one can see that such projection
cannot increase the rank of the OPCM.

At the same time the circumstance that the time evolution of entanglement of specific
particles is determined by the projections of the total OPCM results in non-trivial time
evolution of initially entangled state. As an example let us consider the situation of
small total entanglement, more specifically, when there are only terms with
$\mathfrak{l} = 1$ and $2$ in Eq.~\eqref{eq:u_spdm_spectral} with $\lambda_1
\gg \lambda_2$. According to Eq.~\eqref{eq:Ka_from_Ku} the OPCM for the field
$a$ is given by
\begin{equation}\label{eq:K_a_small}
  \widehat{G}^{(a)} = \lambda^{(a)}_1(t) \mathbf{v}^{(a)}_1(t) \otimes \mathbf{v}^{(a)}_1(t)
  + \lambda^{(a)}_2(t) \mathbf{v}^{(a)}_2(t) \otimes \mathbf{v}^{(a)}_2(t),
\end{equation}
where $\mathbf{v}_{\mathfrak{l}}^{(a)} = \widehat{\Pi}_{a}
\mathbf{v}_{\mathfrak{l}}/\left|\widehat{\Pi}_a
\mathbf{v}_{\mathfrak{l}}\right|$ and $\lambda^{(a)}_{\mathfrak{l}}(t) =
\lambda_{\mathfrak{l}} \left|\widehat{\Pi}_a
\mathbf{v}_{\mathfrak{l}}\right|^2$. Thus, the value of entanglement depends on
the magnitude of the projections $\widehat{\Pi}_a \mathbf{v}_{\mathfrak{l}}$,
which is determined by the internal dynamics of the coupled fields. In particular, if
$\left|\widehat{\Pi}_a \mathbf{v}_{\mathfrak{l}}\right| \ll 1$ one may have
$\lambda^{(a)}_1(t) \sim \lambda^{(a)}_2(t)$ resulting in the significant
entanglement of particles $a$, either with each other or with particles $b$. For more
concrete information one has to take into account that, generally speaking,
Eq.~\eqref{eq:K_a_small} may not be the spectral representation of the matrix
$\widehat{G}^{(a)}$ because the vectors $\mathbf{v}^{(a)}_1(t)$ and
$\mathbf{v}^{(a)}_1(t)$ are not necessarily orthogonal. Let $\theta(t) =
|\mathbf{v}^{(a)}_1(t)\cdot \mathbf{v}^{(a)}_2(t)|^2$, then the eigenvalues of
$\widehat{G}^{(a)}$ are found to be
\begin{equation}\label{eq:egen_K_projected}
  \widetilde{\lambda}^{(a)}_{1,2} = \frac 1 2 \left(\lambda^{(a)}_1 + \lambda^{(a)}_2\right)
  \pm \frac 1 2 \sqrt{\left(\lambda^{(a)}_1 - \lambda^{(a)}_2\right)^2 + 4 \lambda^{(a)}_1\lambda^{(a)}_2 \theta(t)}.
\end{equation}
Thus, if at some particular instant one has $\lambda^{(a)}_1(t) \approx
\lambda^{(a)}_2(t)$ then, depending on the details of the dynamics of the
eigenvectors, one may have $E_N^{(a)} \approx 1$. Strong entanglement is
produced when the ``weak" component of the OPCM $\sim \lambda_2$ is transferred
more effectively than the major component $\sim \lambda_1$, whose only small part
is moved into the field $a$ during the evolution. At the same time, as can be seen
from Eq.~\eqref{eq:egen_K_projected} after slight change of notations, such spike of
entanglement between particles $a$ is accompanied with disentanglement of
particles $b$.

The main condition for this geometric effect is the smallness
of the projection of the respective eigenvector of the total
OPCM. As a result the characteristic feature of the OPCM of
strongly entangled states in this case is
$\mathrm{Tr}[\widehat{G}^{(a)}] \ll \mathrm{Tr}[\widehat{G}] =
\mathrm{Tr}[\widehat{G}^{(a)}] +
\mathrm{Tr}[\widehat{G}^{(b)}]$. That is the states with
$E_N^{(a)} \approx 1$ developed from the states with low total
entanglement are characterized by low excitation, while the
strongly excited field, say the field $b$ in the considered
example, for which one has $\mathrm{Tr}[\widehat{G}^{(b)}]
\approx \mathrm{Tr}[\widehat{G}]$, remains only weakly
entangled.

The fact that  entanglement of the specific particles is determined by the projections
of the total OPCM may lead not only to increased entanglement of the particles but
also to disentanglement. Indeed, if for the OPCM given by
Eq.~\eqref{eq:u_spdm_spectral} with $\mathfrak{l} = 1,2$ at some instant the
vectors $\mathbf{v}_1$ and $\mathbf{v}_2$ belong to different subspaces (say,
$\widehat{\Pi}_a \mathbf{v}_1 = \mathbf{v}_1$ and $\widehat{\Pi}_b
\mathbf{v}_2 = \mathbf{v}_2$) then entanglement of particles of both sorts, $a$
and $b$, are zero. One can understand this effect introducing the isospin quantum
number, so that one value of isospin corresponds to the particle $a$ and another one
stands for the particle $b$. Such situation, when non-zero total entanglement
coexists with zero entanglement as seen from specific particles' OPCM, corresponds to
entanglement stored in isospin.


These effects are of general nature and don't depend on the number of particles and
may play the important role for the dynamics of entanglement in specific systems, for
example, the time dependence of entanglement of initially entangled light in a leaky
cavity.

\section{Entanglement produced by interaction}
\label{sec:bosons}

Simple linear coupling of the quantum fields does not suffice for the fields to become
entangled. The situation, as we show in the present section, is different when there is
an interaction in the system, that is when the energy depends non-linearly on the
number particles. More specifically, we consider the self-interacting two-mode boson
field described by the Hamiltonian
\begin{equation}\label{eq:interaction_boson}
  \HH = \sum_{\kappa=+,-} \epsilon_\kappa a^\dagger_\kappa a_\kappa +
  \sum_{\kappa=+,-} U_\kappa a^\dagger_\kappa a^\dagger_\kappa a_\kappa a_\kappa + U_{+-} a^\dagger_- a^\dagger_+ a_+ a_-.
\end{equation}
Here the first two terms describe the internal dynamics of the modes with
$\epsilon_\kappa$ and $U_\kappa$ being the energies of the modes and the
interaction parameters, respectively, and the last term represents the interaction
between the modes, which is chosen in the form preserving the number of particles within each mode.

Before we apply the general ideology set in Section~\ref{sec:pictures}, we consider
the two-particle case, which provides the clear connection with the standard
quantum-mechanical consideration. In the basis of the population numbers any
two-particle state can be presented as
\begin{equation}\label{eq:two_boson_pop}
  |\psi \rangle = \alpha_{2,0} \ket{2,0} + \alpha_{1,1} \ket{1,1}
  + \alpha_{0,2} \ket{0,2},
\end{equation}
where $\ket{n_+,n_-}$ denotes the state with $n_+$ particles in the ``$+$"-mode
and $n_-$ particles in the ``$-$"-mode, and $\alpha_{n_+,n_-}$ are the respective
amplitudes. Alternatively the state can be presented
as\cite{PASKAUSKAS:2001:ID3811}
\begin{equation}\label{eq:state_creation}
  \ket{\psi} = \sum_{\kappa, \lambda}w_{\kappa, \lambda}a_\kappa^\dagger a_\lambda^\dagger \ket{0},
\end{equation}
where $\ket{0}$ is vacuum and $w_{\kappa, \lambda}$ is a symmetric matrix.
Comparing Eqs.~\eqref{eq:two_boson_pop} and \eqref{eq:state_creation} one finds
$w_{++} = \alpha_{2,0}/\sqrt{2}$, $w_{--} = \alpha_{0,2}/\sqrt{2}$ and
$w_{+-} = \alpha_{1,1}/{2}$. From Eq.~\eqref{eq:state_creation} one obtains the
OPCM
\begin{equation}\label{eq:spdm_boson_int}
  \widehat{G} = 4 \widehat{w}^\dagger \widehat{w}.
\end{equation}

The OPCM is a $2\times2$ matrix and, therefore, its eigenvalues are completely
determined by its determinant $\mathrm{det}[\widehat{G}] = \lambda_1
\lambda_2$ and its trace $\mathrm{Tr}[\widehat{G}] = \lambda_1 + \lambda_2 =
N$ with $N$ being the number of particles. Entanglement, in turn, is determined by
the normalized eigenvalues $\widetilde{\lambda}_{1,2} = \lambda_{1,2}/N$, which
are found as
\begin{equation}\label{eq:normalized_boson}
 \widetilde{\lambda}_{1,2} = \frac 1 2 \pm \frac 1 2 \sqrt{1-\left|\frac{2 C}{N}\right|^2}
\end{equation}
where $|C|^2 = \det{\widehat{G}}$ is the concurrence
\cite{WOOTTERS:1998:ID3357}. In order to see the relation with the standard
definition of the concurrence in the two-particle case we introduce the ``spin flip"
transformation $\sigma_y \ket{+} = -i \ket{-}$, $\sigma_y \ket{-} = i \ket{+}$
and the spin flip state $\ket{\widetilde{\psi}} =
\left(\sigma_y\otimes\sigma_y \ket{\psi}\right)^*$,
then
\begin{equation}\label{eq:concur_psis}
  C = \langle \widetilde{\psi} | \psi \rangle = 4 \det[\widehat{w}].
\end{equation}
Thus the two-particle case completely fits into the canonical
quantum-mechanical description.

In terms of amplitudes $\alpha_{\kappa, \lambda}$ the concurrence is expressed as
\begin{equation}\label{eq:conc_alphas}
  C = 2 \alpha_{2,0}\alpha_{0,2} - \alpha_{1,1}^2.
\end{equation}
For a two-particle state to be disentangled the amplitudes have to meet the
condition $C=0$. Let initially the state be disentangled. It is seen that
this condition not necessarily holds for all $t>0$ if the dynamics of the
system inhomogeneously depends on the population numbers. Since the
Hamiltonian \eqref{eq:interaction_boson} conserves the number of particles
in each mode, the states $\ket{n_+,n_-}$ are the eigenstates and the time
dependence of the amplitudes can be easily found
\begin{equation}\label{eq:inter_amplitudes}
  \alpha_{n_+,n_-}(t) =
     \exp\left[-i t \left(n_+ \epsilon_+ + n_- \epsilon_- + \Delta\epsilon_{n_+,n_-}\right)\right]
     \alpha_{n_+,n_-}(0),
\end{equation}
where $\Delta\epsilon_{n_+,n_-} = U_+ n_+(n_+-1) + U_- n_-(n_--1) +
U_{+-}n_+ n_-$. Substituting Eq.~\eqref{eq:inter_amplitudes} into
Eq.~\eqref{eq:conc_alphas} we find for an initially disentangled state
\begin{equation}\label{eq:c_t_dis}
  C(t) = 2 i \alpha^2_{1,1}(0) e^{-i \Omega t} \sin(\omega t),
\end{equation}
where $\Omega = 2 \epsilon_+ + 2 \epsilon_- + U_+ + U_- + U_{+-}$ and
\begin{equation}\label{eq:ent_beat}
  \omega = U_{+-} - U_+ - U_-
\end{equation}
defines the typical time scale of the entanglement dynamics.

Several interesting conclusions can be drawn from this result. First, entanglement
oscillates between $0$ and the maximum value determined by the contribution of
$\ket{1,1}$ into the initial state. The origin of the oscillations can be traced to the
structure of the concurrence, Eq.~\eqref{eq:conc_alphas}, and the frequencies of the
many-body amplitudes in Eq.~\eqref{eq:inter_amplitudes}. The interaction leads to
the energy shifts $\Delta\epsilon_{nm}$, which depends on the population of the
particular modes. The phase mismatch between the amplitudes results in the
nontrivial time dependence of $|C(t)|^2$.

Second, the interplay between the effects of the intra-mode, $\propto
U_{ii}$, and inter-mode, $U_{+-}$, interactions is not straightforward. Let
the interaction between the modes be absent, $U_{+-} \equiv 0$. As follows
from Eq.~\eqref{eq:c_t_dis}, even in this case initially disentangled states
become entangled.


The related effect is the mutual cancelation of the phase desynchronization
if $U_{+-} = U_{+} + U_{-}$, when despite the interaction, which changes the
energies of the many-body states comparing to multiples of the
single-particle states, initially disentangled states remain disentangled.

The two-particle case is useful for establishing the relation with the standard
description of entanglement. However, in order to grasp the general structure of
entanglement dynamics for the quantum field with interaction it is constructive to
consider more general case with an arbitrary (but definite) number of particles $N$.
This is when the approach based on irreducible representations of the Lie algebra
$\mathfrak{su}(2)$ becomes especially useful. We introduce the operators of
components of angular momentum (see Eq.~\eqref{eq:ang_momentum_ops}) and
enumerate the states by the total angular momentum $j = (n_+ + n_-)/2$ and its
projection $m=(n_+ - n_-)/2$ instead of the population numbers $n_+$ and $n_-$,
so that $\ket{n_+,n_-} =\ket{j,m}_S$. In the following we will omit the index $S$
for brevity.

Expressing the OPCM in terms of the mean values of the operator of the angular
momentum one finds the concurrence $|C(t)|^2 = N^2/4 -
|\aver{\mathbf{J}(t)}|^2$, so that the normalized eigenvalues of the OPCM can be
expressed as
\begin{equation}\label{eq:eigen_SPDM_angular}
\widetilde{\lambda}_{1,2} = \frac 1 2 (1 \pm \widetilde{J}),
\end{equation}
where $\widetilde{J} = 2 |\aver{\mathbf{J}}|/N$. Thus, entanglement can be
written as $E_N = F(\widetilde{J})$, where
\begin{equation}\label{eq:def_F_ent}
  F(x) = \frac 1 2 \sum_{n=1,2} \left[1 + (-1)^n x \right] \log_2\left[1 + (-1)^n x \right].
\end{equation}
In particular,  completely entangled states are characterized by
$|\aver{\mathbf{J}}| = 0$ and disentangled ones are those with
$\left|\aver{\mathbf{J}}\right| = j$. This, of course, agrees with the general result
expressed by Eq.~\eqref{eq:disent_general} (taken for the case of fixed number of
particles). Indeed, $\aver{\mathbf{J}}$ transforms under rotations as a 3d vector,
hence, one can choose a new frame such that $\aver{\quantop{J}_x} =
\aver{\quantop{J}_y} = 0$ and, respectively, $\aver{\quantop{J}_z} = j$. Thus in
this frame the disentangled state has the simple form $\ket{j,j} = (b^\dagger)^N
\ket{0}/ \sqrt{N!}$ (cf. Eq.~\eqref{eq:disent_b}). Conversely, any disentangled
state can be obtained by rotating the state $\ket{j,j}$ in a fixed frame
\begin{equation}\label{eq:disent_inter}
  \ket{\psi(\beta_1,\beta_2)} = \exp(-i \quantop{J}_z \beta_1)
  \exp(-i \quantop{J}_y \beta_2) \ket{j,j},
\end{equation}
where $\beta_i$ are the Euler angles and we have omitted the redundant rotation
around $z$-axis. Thus disentangled states with definite number of particles are
coherent spin states.\cite{PERELOMOV:1986:ID4096,GAZEAU:2009:ID4097}

In terms of the operator of angular momentum
Hamiltonian~\eqref{eq:interaction_boson} can be presented as
\begin{equation}\label{eq:Ham_inter_angular}
  \HH = f_1 + f_2 \quantop{J}_z + \omega \quantop{J}_z^2,
\end{equation}
where $\omega$ is given by Eq.~\eqref{eq:ent_beat} and $f_1 =
\quantop{N}\sum_k\epsilon_k U_k/2 + \quantop{N}^2(U_+ + U_- + U_{+-})/4$
and $f_2 = \epsilon_+ - \epsilon_- + (U_+ - U_-)(1 + \quantop{N})$ depend on the
total number of particles, $\quantop{N} = a_+^\dagger a_+ + a^\dagger_- a_-$,
and are irrelevant for the entanglement dynamics.

As follows from Eqs.~\eqref{eq:eigen_SPDM_angular} and
\eqref{eq:Ham_inter_angular} the dynamics of entanglement is subject to the
general constraint $\aver{\quantop{J}_z(t)} = \aver{\quantop{J}_z(0)}$,
which follows from $[\HH, \quantop{J}_z] = 0$. Thus, the variation of
entanglement is determined by the change with time of the ``transversal"
component $J_\perp^2(t) = \aver{\quantop{J}_x(t)}^2 +
\aver{\quantop{J}_y(t)}^2 = \aver{\quantop{J}_+(t)}\aver{\quantop{J}_-(t)}$,
where $\quantop{J}_\pm(t) = \quantop{J}_x(t) \pm i \quantop{J}_y(t)$. In
particular, this imposes the upper limit on the value of entanglement
produced by the interaction, $E_N(t) \leq E_N^{(max)} =
F[\aver{\quantop{J}_z(0)}/j]$.

\begin{figure}
  \includegraphics[width=5in]{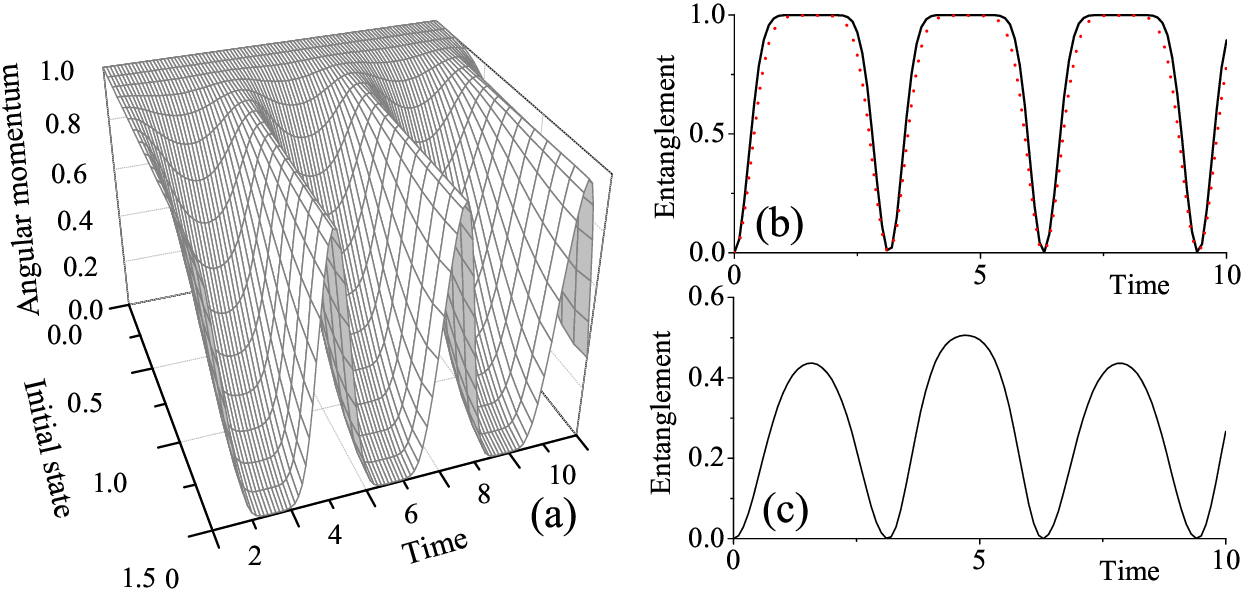}\\
  \caption{Dependence of entanglement of of the two-mode self-interacting boson field model on time and
  on structure of the initial disentangled four-particle state. (a) Normalized angular momentum $\widetilde{J}(\beta,t)$
 as a function of time (in units $\omega^{-1}$) and angle $\beta$ characterizing the initial state. (b) Time dependence of linear entropy
 $E_L = 2(1- \mathrm{Tr}[\widehat{\rho}^2]) $ (solid line) and von Neumann entropy
 $E_N = - \mathrm{Tr}[\widehat{\rho} \log_2 \widehat{\rho}]$ (dotted line) for
 $\beta = \pi/2$ (when $\aver{\quantop{J}_z} = 0$). Different measures of entanglement follow very close each other. (c) Time dependence
 of entanglement $E_N(t)$ for $\beta = \pi/4$.
  }\label{fig:jinter}
\end{figure}

Let us consider the dynamics of initially disentangled states. Rotations around the
$z$-axis do not affect $J_\perp^2$ and, therefore, the dynamics of entanglement of
the states related through such rotations are identical. Hence, it suffices to consider
only states $\ket{\psi(\beta)} = \ket{\psi(0,\beta)}$. From $[\HH,
\quantop{J}_z] = 0$ it follows that
\begin{equation}\label{eq:jz_inter}
  \aver{\quantop{J}_z(t)}_\beta = j \cos(\beta),
\end{equation}
where the index $\beta$ emphasizes the structure of the initial state, i.e.
$\aver{\ldots}_\beta = \bra{\psi(\beta)} \ldots \ket{\psi(\beta)}$. The nontrivial
part of the time dependence of $\aver{\quantop{J}_\pm(t)}_\beta$ is given only by
the last term in Eq.~\eqref{eq:Ham_inter_angular} because the first two terms yield
only the phase factor, which does not contribute to $\widetilde{J}^2(t)$. Thus we
can simplify the discussion considering $\HH_0 = \omega \quantop{J}_z^2$ instead
of full Eq.~\eqref{eq:Ham_inter_angular}. Next, we notice that the solutions of the
operator equations of motion $\dot{\quantop{J}}_\pm(t) = i[\HH_0,
{\quantop{J}}_\pm(t)]$ have the form
\begin{equation}\label{eq:jxy_sols}
  \quantop{J}_\pm(t) = \exp\left({-i \omega t \pm 2i \omega \quantop{J}_z t}\right) \quantop{J}_\pm.
\end{equation}
Using these solutions we obtain (see Appendix~\ref{sec:typical_average} for
the details)
\begin{equation}\label{eq:J_p_explicit}
\begin{split}
  \widetilde{J}^2 = & \cos^2(\beta) +  \\
  & + 4 \sin^2(\beta/2)
  \cos^{2N - 2}(\theta/2) \\
  & \times \left| e^{i \gamma} \sin(\beta/2)\sin(\theta/2) + \cos(\beta/2)\cos(\theta/2)\right|^2,
\end{split}
\end{equation}
where $\sin(\theta/2) = \sin(\beta) \sin(\omega t)$ and $\cot(\gamma) = \cos(
\beta) \tan(\omega t)$. The overall dependence of $\widetilde{J}$ on time and on
the structure of the initial state parametrized by the angle $\beta$ is shown in
Fig.~\ref{fig:jinter}. We would like to note that the normalized angular momentum
$\widetilde{J}$ determines the linear entropy $E_L = 2(1 -
\mathrm{Tr}[\widehat{\rho}^2])$ in a simple way $E_L = 1 - \widetilde{J}^2$. In
turn, the overall profiles of $E_L(t)$ and $E_N(t)$ are very similar, as illustrated by
Fig.~\ref{fig:jinter}b, and, therefore, one can infer the general features of
entanglement directly from $\widetilde{J}(t)$.

It follows from Eq.~\eqref{eq:J_p_explicit} that for $\beta \ne 0$ or $\pi$ the
transverse component of the effective angular momentum oscillates with the period
$T=2\pi/\omega$. The maxima of $\widetilde{J}$, which, according to
Eq.~\eqref{eq:eigen_SPDM_angular}, correspond to the minima of entanglement,
are reached at $t_{min} = T n/2$ with integer $n$. As follows from
Eq.~\eqref{eq:J_p_explicit}, $\widetilde{J}(t_{min}) = 1$ yielding $E_N = 0$ in
agreement with the definition of the states $\ket{\psi(\beta)}$. There are two sets of
maxima of entanglement (see Fig.~\ref{fig:jinter}c) at $t_{max}^{(1)} = (1/4 +
n)T$ and $t_{max}^{(2)} = (3/4 + n)T$, where one has
\begin{equation}\label{eq:j_min}
  \begin{split}
  \widetilde{J}^2(t_{max}^{(1)}) = \cos^2(\beta)+ \sin^2(\beta)\cos^{2N - 2}(\beta), \\
 \widetilde{J}^2(t_{max}^{(2)}) = \cos^2(\beta)+ 4 \sin^2(\beta/2)\cos^2(3\beta/2)\cos^{2N - 2}(\beta),
  \end{split}
\end{equation}
respectively. One can see that $\widetilde{J}^2(t_{max}^{(1)})/
\widetilde{J}^2(t_{max}^{(2)}) \geq 1$. With increasing the number of
particles the ratio monotonously tends to $1$.

In order to  better understand the effect of the number of particles on the time
evolution of entanglement we consider the case $\beta = \pi/2$, that is the case of
initially disentangled states with symmetrically populated modes ($\bra{\psi}
\quantop{J}_z \ket{\psi} = 0$). Up to rotation around $z$-axis these are the only
states (among initially disentangled ones) that yield maximum entanglement $E_N =
1$ in the course of time evolution.
From Eq.~\eqref{eq:J_p_explicit} one finds that for these states
\begin{equation}\label{eq:jpmax}
  \widetilde{J}^2 = \cos^{2N - 2} (\omega t).
\end{equation}
The time dependence of entanglement following from Eq.~\eqref{eq:jpmax} has two
specific features. First, this is the periodic function of time. Entanglement
considerably changes (practically from $1$ to $0$ and back) within the vicinities
$t_{min}$. When the number of particles increases these regions narrow. It is
constructive to consider the limiting form of the time dependence when the number
of particles becomes very large. In the limit $N \gg 1$ one can approximate (see
Appendix~\ref{sec:typical_average})
\begin{equation}\label{eq:jpmax_Gauss}
  \widetilde{J}^2(t) \approx \sum_{n = 0}^\infty \exp\left[ - (N - 1)(\omega t - \pi n)^2\right].
\end{equation}
Thus, away from the points where $\widetilde{J}(t) = 0$, it can be regarded as the
sequence of the Gaussian bumps localized near $t_{min}$. As a result, when
$\sqrt{N} \gg 1$ one can consider the system as spending most of the time in states
with high entanglement, $E_N \approx 1$.

The Gaussian decay of $\widetilde{J}(t)$ is similar to the Gaussian decay of
coherence of central system, \cite{MELIKIDZE:2004:ID3941,LAGES:2005:ID3940}
two spins $1/2$ coupled to the bath. In particular, the same dependence of
the decay rate on the number of particles $\propto \sqrt{N}$ in the
environment [notice $N-1$ in Eq.~\eqref{eq:jpmax_Gauss}] should be
emphasized. There are, however, two important differences between this
situation and our case. First, the Gaussian decay for the case of
central system appears in the limit of slow dynamics of the environment.
In the opposite limit the decay follows the Lorentsian law and in the
intermediate case both types of decays present at different time scales.
\cite{YUAN:2009:ID3939} For the two-mode boson field it is meaningless to
separate particular particles and environment due to indistinguishability,
however, it is worth noting that $\widetilde{J}(t)$ does not depend on the
single-particle energies. The second important difference is that the
concurrence of the two-mode boson field exhibits oscillations while the loss
of coherence of the central system is irreversible.

Despite the complex structure of the whole manifold of completely entangled states,
the maximal entangled states reached in the course of evolution described by the
Hamiltonian $\HH_0 = \omega \quantop{J}_z^2$ are simple and belong to the class
of NOON-states. The maximal entanglement is reached for the first time at $\omega t
= \pi/2$, when the state of the system can be presented as
\begin{equation}\label{eq:ent_st-max}
  \ket{\chi}= \sum_{m=-j}^{j}
            e^{-i m^2 \pi/2} \ket{j,m} d_{m,j}(\pi/2),
\end{equation}
where $d_{m,j}(\theta) = \bra{j,m} \exp(-i \quantop{J}_y \theta) \ket{j,j}$.

First we consider the case of integer $j$. The amplitudes $\chi_m = e^{-i m^2
\pi/2}$ are $1$ and $-i$ for even and odd $m$, respectively, and can be presented
as
\begin{equation}\label{eq:amplitudes_NOON}
  \chi_m = \frac{1}{\sqrt{2}}\left(e^{-i\pi/4} + (-1)^m e^{i\pi/4}\right).
\end{equation}
Next, we write $(-1)^m = i^{2j} (-1)^{m-j}$ and use the
symmetry $d_{m,-j}(\pi/2) = (-1)^{m-j}d_{m,j}(\pi/2)$. Now we
can sum over $m$ and obtain
\begin{equation}\label{eq:NOON_integer}
 \ket{\chi} = e^{-i \quantop{J}_y \pi/2} e^{-i \quantop{J}_z \pi/(4j)} \frac{1}{\sqrt{2}} \left[\ket{j,j}
 + i^{2j} \ket{j,-j}\right].
\end{equation}
These are NOON-states as they were defined in
Eq.~\eqref{eq:NOON_def}: the basis vectors spanning the state
are obtained by rotating $\mathbf{S}(1)=(1,0)$ and
$\mathbf{S}(2)=(0,i)$.

For the case of half-integer $j$ we introduce $\widetilde{m} =
m - 1/2$ so that $e^{-i m^2 \pi/2} = \exp(-i\pi/8 - i\pi m/2 -
i \pi \widetilde{m}^2/2)$.  Now the same arguments as above can
be repeated leading to
\begin{equation}\label{eq:NOON_half-integer}
 \ket{\chi} = e^{-i \quantop{J}_z \pi/2} e^{-i \quantop{J}_y \pi/2} e^{-i \quantop{J}_z \pi/(4j)}
 \frac{e^{i\pi/8}}{\sqrt{2}} \left[\ket{j,j}
 + i^{2j-1} \ket{j,-j}\right].
\end{equation}


So far in this section we have considered the case when states of the system are
characterized by a definite number of particles. In order to give a complete
description how entanglement is produced for initially disentangled states it is
constructive to include into consideration states that are initially disentangled within
both, field and particle, pictures. As has been shown in Section~\ref{sec:pictures},
these are (Glauber's) coherent states and in accord with
Eq.~\eqref{eq:disent_general} they can be presented as
\begin{equation}\label{eq:disent_both}
  \ket{\psi(t=0)} = e^{-|\alpha|^2/2} \sum_{N=0}^\infty \frac{\alpha^N}{\sqrt{N!}}
  \ket{\psi_N(\mathbf{S})},
\end{equation}
where $\alpha$ is a complex number and $\ket{\psi_N(\mathbf{S})}$ are states
with $N=2j$ particles in the same state characterized by vector $\mathbf{S}$.
Following Eq.~\eqref{eq:disent_inter} all such states are obtained by the same
rotations of respective $\ket{j,j}$ states.

Since operators of angular momentum $\quantop{J}_{x,y,z}$
preserve the total number of particles the average values
$\aver{\quantop{J}_i}$ are simply found as weighted sum of
contributions from components with definite number of particles
\begin{equation}\label{eq:weighted_sum}
  \aver{\quantop{J}_i}= e^{-|\alpha|^2} \sum_N \frac{|\alpha|^{2N}}{N!} \aver{\quantop{J}_i(t)}_N,
\end{equation}
where $\aver{\quantop{J}_i(t)}_N$ is the average taken within the $N$-particle
``sector" --- $\aver{\quantop{J}_i(t)}_N = \bra{\psi_N(\mathbf{S})}
\quantop{J}_i(t)\ket{\psi_N(\mathbf{S})}$.

As well as before the maximum value of entanglement can be reached if the initial
state satisfies $\aver{\quantop{J}_z} = 0$, that is if $\ket{\psi_N(\mathbf{S})}$ is
obtained by rotating around $y$ axis by $\pi/2$. For this case we obtain
\begin{equation}\label{eq:jp_coh}
  \aver{\quantop{J}_+(t)} = \frac{|\alpha|^2}{2}\exp\left[-i\omega t - 2 |\alpha|^2 \sin^2(\omega t/2)\right]
\end{equation}
and $\aver{\quantop{J}_-(t)} = \aver{\quantop{J}_+(t)}^*$. Taking into account
that the average number of particles is related to parameter $\alpha$ by $\aver{N} =
|\alpha|^2$ we obtain for the normalized angular momentum
$\widetilde{J} = 2 |\aver{\quantop{J}_+}|/\aver{N}$
\begin{equation}\label{eq:jtild_coh}
  \widetilde{J}(t) = \exp\left[-2 \aver{N} \sin^2(\omega t/2)\right].
\end{equation}
Similarly to the case considered before $\widetilde{J}(t)$ for $\aver{N} \gg 1$ has a
form of a periodic sequence of Gaussians centered at $t = 2 \pi n/\omega$ with
integer $n$. The significant difference is that owing to contribution of one-particle
(obviously disentangled within the particle picture) states, the smallest value reached
by $\widetilde{J}(t)$ is $\widetilde{J}_{min} = \exp(-2\aver{N}) \ne 0$ implying
formal absence of complete entanglement. At the same time it should be noted that
when $\widetilde{J}(t) = \widetilde{J}_{min}$ all many-particle states are
NOON-states.

\section{Entanglement dynamics in Jaynes-Cummings model}
\label{sec:jc-model}

In the previous sections we have seen that the entanglement dynamics in the system
of coupled quantum fields is characterized by two specific features --- redistribution
of the initial total entanglement and the production of entanglement by
self-interaction. In the present section we apply these ideas for the analysis of
entanglement in a more complex situation.

We consider a single two-level atom interacting with the quantized
electromagnetic field. The electron transitions in the atom are assumed to
be characterized by definite helicity. The excitation of the electron state
with the spin down at the ground level,  $\ket{g\downarrow}$, into the spin
up state at the excited level $\ket{e\uparrow}$ occurs through the
absorption of ``$+$"-polarized photon and so on. The dynamics of the system
is described by the Hamiltonian of the two-mode Jaynes-Cummings (JC)
model,\cite{SHORE:1993:ID3860} which we write down in terms of the creation
and annihilation operators
\begin{equation}\label{eq:jc_two-mode}
\HH = \sum_\kappa \epsilon^{(p)}_\kappa a_\kappa^\dagger a_\kappa
   + \sum_k \epsilon^{(e)}_k c_k^\dagger c_k + \HH_+ + \HH_-.
\end{equation}
Here the first two terms describe the dynamics of the free atom and the free field,
respectively, with $k$ and $\kappa$ running over the atomic states, $\{g\uparrow,
g\downarrow, e\uparrow, e\downarrow\}$, and the photon polarizations, $+$ and
$-$, respectively. The interaction between the atom and photons is described by
$\HH_\kappa = \omega_\kappa a_\kappa \sigma^\dagger_\kappa +
\mathrm{h.c.}$, where $\sigma^\dagger_+ = c_{e\uparrow}^\dagger
c_{g\downarrow}$, $\sigma^\dagger_- = c_{e\downarrow}^\dagger
c_{g\uparrow}$ and $\omega_\pm$ are the respective Rabi frequencies.

The photon entanglement is defined as the von Neumann entropy of the (properly
normalized) single-photon correlation matrix with the matrix elements
\begin{equation}\label{eq:photon_DM}
  {G}_{\kappa, \lambda}(t) = \aver{a_\kappa^\dagger(t) a_{\lambda}(t)}.
\end{equation}
This is a $2$ by $2$ matrix and the problem of entanglement can be approached
using the same description as in the previous section. As well as before our main
objective is to study the time evolution of entanglement of initially disentangled
photonic states. This closely corresponds to the situation when, for example, the
cavity in the ground state is pumped by an external source.

Taking into account that initially the atom is not excited the initial state of the system
is presented as
\begin{equation}\label{eq:jc_initial_state}
  \ket{\Psi(0)} = \ket{\psi(\beta)} \ket{0}_e,
\end{equation}
where $\ket{\psi(\beta)}$ is a disentangled photon state obtained by rotations of
$\ket{j,j}_S$ [see Eq.~\eqref{eq:disent_inter}] and $\ket{0}_e$ denotes the state
of the atom with both electrons at the ground level.

The time dependence of photon entanglement of few-photon states is highly
nontrivial comparing to the few particle states of the two-mode boson field considered
in the previous section. Here we would like to note only one characteristic feature.
Drawing an analogy with the consideration of the two-particle case in the previous
section one can expect that in the present case as well the time dependence of
entanglement will be determined by the mismatch between the amplitudes of states
with different population numbers. In fact, as will be evident shortly, the typical
frequencies are determined by \textit{the square roots} of population numbers.
Taking this circumstance into consideration one can expect that the time dependence
of the concurrence is the result of superposition of several harmonics with
incommensurate frequencies. Thus the concurrence is a quasi-periodic function with a
complex profile.\cite{EREMENTCHOUK:2010:ID4101}

\begin{figure}
  \includegraphics[width=4in]{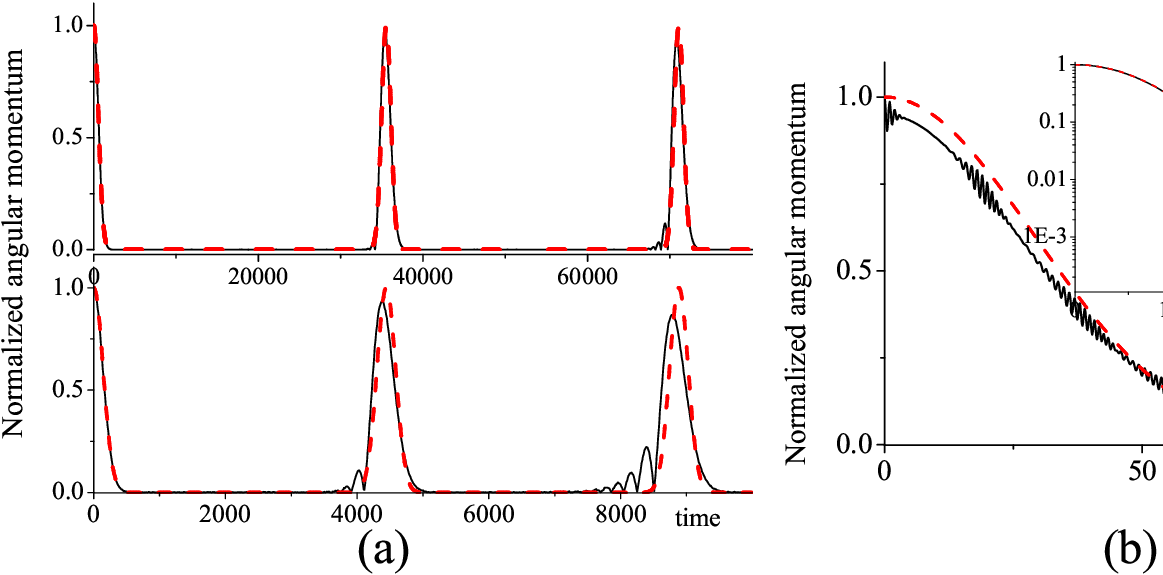}\\
  \caption{Time dependence of $\widetilde{J}(t)$ for the Jaynes-Cummings model. The
  time is measured in inversed coupling constant $\omega_R$.
  (a) Comparison of the exact dynamics (dotted lines) with the approximation defined by
  Eq.~\eqref{eq:effective_H_final} (dashed lines) on the long time scale
  for the system with $50$ (lower panel) and
  $200$ (upper panel) photons.
  (b) The short time dependence for $10$ particles --- the exact result (solid line)
  and the approximation (dashed line).
  The transition to the Gaussian profile occurs fairly quickly with the number of
  particles. It is illustrated  in inset, where the short time
  dependence is shown for $200$ photons in log-scale.}\label{fig:jc_j}
\end{figure}

With increasing the number of photons, however, the contribution of specific
frequencies becomes less important and the overall shape of $\widetilde{J}(t)$
changes toward some general regular pattern as illustrated in
Fig.~\ref{fig:jc_j}. In order to describe it we use the explicit form of Heisenberg
representation for the photon operators $a_\kappa(t) = \exp(i t
\quantop{H}_\kappa) a_\kappa \exp(-i t \quantop{H}_\kappa)$. Taking into
account the separability of polarizations dynamics we have\cite{Scully_QO}
\begin{equation}\label{eq:ev_op}
  a_\kappa(t) = e^{-i\epsilon^{(p)}_\kappa t + i \quantop{C}_\kappa t} \left\{
    \left[\cos\left(\bar{\quantop{C}}_\kappa t\right) - i \quantop{C}_\kappa
    \frac{\sin\left(\bar{\quantop{C}}_\kappa t\right)}{\bar{\quantop{C}}_\kappa}
    \right] a_\kappa - i \frac {\sin\left(\bar{\quantop{C}}_\kappa t\right)}{\bar{\quantop{C}}_\kappa} \sigma_\kappa
  \right\},
\end{equation}
where $\quantop{C}_\kappa^2 = \frac{1}{4}\delta_\kappa^2 + \omega^2_R
a_\kappa^\dagger a_\kappa$ and $\bar{\quantop{C}}_\kappa^2 =
\quantop{C}_\kappa^2 + \omega^2_R $ with $\delta_+ =
\epsilon^{(e)}_{e\uparrow} - \epsilon^{(e)}_{g\downarrow} - \epsilon^{(p)}_+$
and $\delta_- = \epsilon^{(e)}_{e\downarrow} - \epsilon^{(e)}_{g\uparrow}  -
\epsilon^{(p)}_+$ being the detunings from the resonances for $+$- and
$-$-polarized transitions, respectively. It should be noted that the second term in
Eq.~\eqref{eq:ev_op} does not contribute to $G_{\kappa, \lambda}(t)$ because it
vanishes while acting on the initial (ground) state of the atom.

As the consideration in the previous sections suggests, the main effect on the
entanglement dynamics is due to the coupling between the systems rather than due
to their internal dynamics. Therefore, we first consider the simplest resonant case
when one may set $\epsilon^{(e)}_k = 0$ and $\epsilon^{(p)}_\kappa = 0$.
Additionally we assume that the symmetry between transitions with different
helicities is not broken so that $\omega_+ = \omega_- = \omega_R$.

Due to the absorption by the atom of a single photon of either polarization
$\aver{\quantop{J}_z(t)}$ varies with time in contrast to what we had for the
two-mode boson model. This variation, however, is small in the limit of large number
of photons since the interaction changes the mismatch between $+$- and
$-$-polarized photons by $\pm 1$ at most. Indeed, taking into account the identity
$f\left(\bar{\quantop{C}}_\kappa\right) a_ \kappa = a_\kappa
f\left({\quantop{C}}_\kappa\right)$, which holds for any well-behaving function $f$,
we find
\begin{equation}\label{eq:jc_Jz}
 \aver{\quantop{J}_z(t)} = \aver{\quantop{J}_z} +
    \frac 1 2 \aver{\sin[(\quantop{C}_- - \quantop{C}_+) t]\sin[(\quantop{C}_- + \quantop{C}_+) t]},
\end{equation}
where the average is taken over the initial photons state, $\aver{\ldots} =
\bra{\psi(\beta)} \ldots \ket{\psi(\beta)}$. In the case when all photons have the
same polarization the interaction produces small entanglement $\sim 1/N$ oscillating
with the frequency $N\omega_R$. In the unpolarized case, when
$\aver{\quantop{J}_z}=0$, the last term in Eq.~\eqref{eq:jc_Jz} vanishes
identically because of the symmetry of such disentangled states with respect to
flipping all photon spins. In the weakly polarized case the ratio of the last term to
$\aver{\quantop{J}_z}$ remains limited from above and decreases with the number
of particles $\propto 1/\sqrt{N}$.

Thus, while $\aver{\quantop{J}_z(t)}$ is not a constant its total variation is small in
the limit $N \gg 1$. Therefore, the initial value $\aver{\quantop{J}_z(0)}$ can be
considered as imposing a limitation on the highest entanglement, which can be
reached. Because of these reasons, we limit ourselves to considering in details
the case when $\aver{\quantop{J}_z(0)} = 0$, that is when the initial photon state
is given by Eq.~\eqref{eq:disent_inter} with $\beta_2 = \pi/2$. In this case the
eigenvalues of the photon OPCM, and, hence, entanglement, are determined by the
magnitude of $\aver{\quantop{J}_+(t)}$. We evaluate it assuming that the main
contribution is due to the states with small $m$, that is due to $n_+ \approx n_-$
while $n_+, n_- \gg 1$. This approximation improves with increasing the number of
particles.

The best way to employ this approximation is to derive the effective photon
Hamiltonian directly from Eq.~\eqref{eq:ev_op}. Taking into account that initially the
atom is in the ground state we can effectively replace Eq.~\eqref{eq:ev_op} up to
terms $\propto 1/\sqrt{N}$ by $a_k(t) = \exp(i \quantop{C}_k t) a_k \exp(-i
\quantop{C}_k t)$, which is the Heisenberg representation induced by the
Hamiltonian
\begin{equation}\label{eq:effective_H}
  \quantop{H} = \omega_R \sqrt{j + \quantop{J}_z} + \omega_R \sqrt{j - \quantop{J}_z}.
\end{equation}
Using the assumption regarding the main contribution due to small $m$
we expand this Hamiltonian with respect to $\quantop{J}_z/j$ finding
in the first nonvanishing order
\begin{equation}\label{eq:effective_H_final}
  \quantop{H} = \omega_R \sqrt{j} - \frac{\omega_R}{4 j^{3/2}} \quantop{J}_z^2.
\end{equation}
Comparing this expression with Eq.~\eqref{eq:Ham_inter_angular} we come to the
conclusion that in this limit the photons behave as the two-mode self-interacting
boson field with the effective interaction strength $\omega = \omega_R/4 j^{3/2}$.
The characteristic feature of the effective interaction is that its intensity decreases
with the number of particles $\propto 1/N^{3/2}$. This ``spreading" is caused by
sharing the interaction with the single atom among all photons and leads to different
dynamics of $\widetilde{J}(t)$ comparing to what we have seen in the previous
section. Using the results obtained there we find the period of the long-scale
oscillations $T = \sqrt{2 N^3}/\omega_R$ and the decay time (or the entanglement
time) $\tau = \sqrt{2}N/\omega_R$ (see Fig.~\ref{fig:jc_j}a).

It should be emphasized that entanglement obtained here is a result of the effective
photon-photon interaction and cannot be reduced to entanglement in the
photon-atom system or an entanglement transfer (initially the atom is in the ground
state and the state of the whole photon-atom system is separable). Entanglement is
produced in the course of multiple photons absorptions and re-emissions and,
thereby, is in striking contrast with the case when only few such processes
occur.\cite{CAVALCANTI:2006:ID4098}

Fig.~\ref{fig:jc_j} demonstrates that as we go toward longer
times the shape of the bumps changes
--- they become asymmetrical and acquire the side oscillations, the revival
of coherence occurs non-monotonously. The approximation we have used misses
these long-time changes, which are the reminiscent features of the quasi-periodicity
mentioned above. If, however, one restricts to the initial growth of entanglement (the
drop of coherence) these features are not important, and one has in the limit $N\gg1$
a simple expression for initial decay of the normalized angular momentum
\begin{equation}\label{eq:jc_jp_limit}
  \widetilde{J}(t) = e^{-(\omega_R t/2 N)^2}.
\end{equation}
From the formal point of view the approximation leading to effective
Hamiltonian~\eqref{eq:effective_H} emerges in the limit when the relative variation
of parameters $\quantop{C}_\pm$ in Eq.~\eqref{eq:ev_op} with the number of
photons is small. The significant simplification of dynamics in this limit takes place
also for the general model~\eqref{eq:jc_two-mode}, not necessarily in the resonant
regime. In this case the effective Hamiltonian has the form
\begin{equation}\label{eq:effective_H_general}
  \widetilde{\quantop{H}} = \sqrt{\frac{1}{4} \delta_+^2 + \omega_+^2(j + \quantop{J}_z)}
  + \sqrt{\frac{1}{4} \delta_-^2 + \omega_-^2(j - \quantop{J}_z)},
\end{equation}
For initially disentangled states with $\aver{\quantop{J}_z(0)} \approx 0$ this
expression may be expanded up to terms quadratic in $\quantop{J}_z$ yielding a
Hamiltonian of the form~\eqref{eq:Ham_inter_angular}. As has been noted in the
previous section, the constant and linear in $\quantop{J}_z$ terms do not affect the
time dependence of $\widetilde{J}(t)$, which is determined by the effective
interaction parameter (the parameter in front of $\quantop{J}_z^2$)
\begin{equation}\label{eq:omega_detuned}
  \omega = -\frac{\omega_+}{8\left(j+\frac{\delta_+^2}{4\omega_+^2}\right)^{3/2}} -
        \frac{\omega_-}{8\left(j + \frac{\delta_-^2}{4\omega_-^2}\right)^{3/2}}.
\end{equation}
Comparing with Eq.~\eqref{eq:effective_H_final} one finds that in the case of
symmetry between $+$- and $-$-polarized transitions, $\delta_\pm = \delta$ and
$\omega_\pm = \omega_R$, the effect of detuning from the resonance on the
entanglement dynamics can be accounted by substitution $j \to j  +
\delta^2/4\omega_R^2$, which shows that detuning from the resonance relaxes the
requirement imposed on the number of photons.

\section{Conclusion}

We have considered the basic dynamics of entanglement in the system of coupled
second quantized fields in the context of the problem of solid based sources of
entangled light. This compels to treat entanglement as a property of particles, which
are excitations of respective fields, rather than a property of states of fields
themselves. The properties of particles and fields are described by different
quantities: many-particle density matrix and fields density matrix. We show that
while these quantities describe the same state of the physical system they yield
different entanglement. This reflects the fact that entanglement depends on the
notion of locality: \textit{what} is the part of the system and what is the
complement. These notions are clearly different whether we address properties of
particles or fields, in other words whether we consider the system within the particle
or field picture. We show that the same state may be completely entangled in one
picture and disentangled in another. The class of states that are completely
disentangled in both pictures is very simple: these are either states when only
excitations of one type are present or (Glauber's) coherent states. Taking these
considerations into account we paid the most attention to the dynamics of
entanglement in the particle picture, which is the most relevant for the problem of
solid based sources of entangled light.

The first question, which has to be answered is how is it possible to produce an
entangled state of a quantum field. The circumstance, which motivates this question,
is that by a classical source entangled states (in either picture) cannot be reached out
of vacuum. Moreover, we show that in a system of linearly coupled quantized fields
total entanglement (in particle picture) conserves and its dynamics reduces to mere
transfer between subsystems.

The simplest system demonstrating entanglement of initially disentangled states is
the two-mode boson field with self-interaction. Reformulating the problem using the
formalism of Schwinger's model of angular momentum we show that there is the
direct relation between $J$, the magnitude of the average angular momentum
$\aver{\mathbf{J}}$, and the concurrence. More specifically, the states with
maximum possible magnitude of the average angular momentum, or \textit{spin}
coherent states, are disentangled and those with $J=0$ are completely entangled,
meaning that the one-particle correlation matrix is proportional to the identity
matrix.

We show that in the limit of large number of particles $J(t)$  has the overall form of
the periodic sequence of Gaussian bumps, whose width (inverse entanglement time)
decreases with the number of particles $\propto 1/\sqrt{N}$. The interesting feature
is that entanglement is produced even if the interaction exists between the particles
within the same mode but not between different modes. This phenomenon is in the
striking contrast with the standard quantum mechanical picture, where separable
dynamics cannot entangle initially disentangled particles. The origin of this effect lies
in the structure of many-body states. Because of indistinguishability of the particles
one cannot say which particle belongs to which mode. As a result only such states are
disentangled whose amplitudes meet the special condition, which can be broken if the
dynamics nonlinearly depends on the number of particles.

We apply these results for analysis of the two-mode Jaynes-Cummings model.
The photon-atom interaction has been assumed to be helicity preserving thus
the dynamics of ``$+$"- and ``$-$"-polarized photons are separable. However,
absorption and re-emission of photons by the atom introduces an effective
interaction between the photons of the same polarizations resulting in
photon entanglement. This situation provides the clear illustration of the
physical origin of entanglement for the field with apparently separable
dynamics (or, say, the two-mode boson field with absent inter-mode
interaction). Due to indistinguishability all photons are always in the
superposition of states with different polarizations (except, of course,
when the system is completely polarized) and, therefore, is always affected
by ``both parts" of the dynamics.

We show that the effective interaction leads to the typical Gaussian drop of
coherence in the limit of large number of photons $N\gg1$. Since the
interaction with the single atom is shared among the photons the strength of
the effective interaction drops $\propto 1/N^{3/2}$. This leads to prolonged
both the oscillations of entanglement $T = \sqrt{2 N^3}/\omega_R$ and the
entanglement time $\tau = N \sqrt{2}/\omega_R$.

\acknowledgments

We acknowledge support from NSF Grant No. ECCS-0725514, DARPA/MTO Grant No.
HR0011-08-1-0059, NSF Grant No. ECCS-0901784, AFOSR Grant No.
FA9550-09-1-0450, and NSF Grant No. ECCS-1128597.

\appendix

\section{Evolution of the transversal angular momentum}
\label{sec:typical_average}

For different situations considered in the main text we need the value of the typical
matrix element
\begin{equation}\label{eq:typical_average}
  J_+(\phi,\beta) = \bra{\psi(\beta)} \exp(i\quantop{J}_z \phi) \quantop{J}_+ \ket{\psi(\beta)},
\end{equation}
where $\ket{\psi(\beta)} = \exp(-i \quantop{J}_y \beta)\ket{j,j}_S$.

Using the transformation rule
\begin{equation}\label{eq:transf_plus}
  e^{i\quantop{J}_y \beta}\quantop{J}_+ e^{-i\quantop{J}_y \beta} = \quantop{J}_z \sin(\beta)
  + \quantop{J}_+ \cos^2(\beta/2) - \quantop{J}_-\sin^2(\beta/2),
\end{equation}
we can rewrite Eq.~\eqref{eq:typical_average} as
\begin{equation}\label{eq:typical_step1}
  J_+(\phi,\beta) = {}_S \bra{j,j} {\quantop{R}} [\quantop{J}_z \sin(\beta) -
   \quantop{J}_-\sin^2(\beta/2)]\ket{j,j}_S,
\end{equation}
where we have taken into account that $\quantop{J}_+\ket{j,j}_S \equiv 0$
and have introduced ${\quantop{R}} = \exp(i\quantop{J}_y \beta)
\exp(i\quantop{J}_z \phi)\exp(-i\quantop{J}_y \beta)$ the operator of
rotation by angle $\phi$ around the axis along the direction
$(\sin(\beta),0, \cos(\beta))$. Expanding ${\quantop{R}}$ in terms of the
Euler angles $(\gamma',\theta,\gamma)$ related to $\beta$ and $\phi$ through
\begin{equation}\label{eq:Euler_angles}
  \begin{split}
  \sin(\theta/2) & = \sin(\beta) \sin(\phi/2), \\
  \tan[(\gamma'+\gamma)/2] & = \cos(\beta)\tan(\phi/2), \\
  \gamma - \gamma' & = \pi,
  \end{split}
\end{equation}
we find
\begin{equation}\label{eq:jplus_answer}
  J_+(\phi,\beta) = e^{-i(\gamma' + \gamma)} \left[
  j \sin(\beta) d_{j,j}(\theta) - e^{i\gamma}\sqrt{2j} \sin^2(\beta/2) d_{j-1,j}(\theta)
    \right].
\end{equation}
Here
\begin{equation}\label{eq:matrix_rotation}
  d_{m,j}(\theta) = \sqrt{\frac{(2j)!}{(j+m)!(j-m)!}} \cos^{j+m}(\theta/2) \sin^{j-m}(\theta/2)
\end{equation}
are the matrix elements between $\ket{j,j}_S$ and $\ket{j,m}_S$ of the
$(2j+1)$-dimensional irreducible representation of $\exp(-i\quantop{J}_y
\theta)$.\cite{SAKURAI:1994:ID3942} Using Eq.~\eqref{eq:matrix_rotation} in
Eq.~\eqref{eq:jplus_answer} we finally obtain
\begin{equation}\label{eq:jplus_final}
  J_+(\phi,\beta) = 2j e^{-i(\gamma' + \gamma)}\sin(\beta/2) \cos^{2j-1}(\theta/2)\left[\cos(\beta/2)\cos(\theta/2)
  -e^{i\gamma}\sin(\beta/2)\sin(\theta/2)\right].
\end{equation}
In the case of the special interest $\beta = \pi/2$ this expression
significantly simplifies
\begin{equation}\label{eq:jplus_t}
 J_+(\phi,\pi/2) = j e^{-i \phi/2}\cos^{2j-1}(\phi/2).
\end{equation}

For small $\phi$ and large $j$ the magnitude of this function decays with $\phi$
following the Gaussian law. Let us denote $\widetilde{J}(\phi) =
\cos^{2j-1}(\phi/2)$, then
\begin{equation}\label{eq:dft}
  \frac{d \widetilde{J}(\phi)}{ d\phi} = \frac 1 2 (2j-1) \tan(\phi/2) \widetilde{J}(\phi).
\end{equation}
For $\phi \ll 1$ we may keep only the leading term in the Taylor expansion of
$\tan(\phi/2)$ and obtain
\begin{equation}\label{eq:ft_gauss}
  \widetilde{J}(\phi) \approx \exp\left[-\frac 1 2 (2j -1) \left(\frac{\phi}{2}\right)^2\right].
\end{equation}
Such representation is meaningful for sufficiently large $j$ when $j\phi^2
> 1$. The generalization of this representation for $\phi \approx 2 \pi n$ with integer $n$, where $\sin(\phi/2) \approx
0$, is obvious.


\end{document}